\documentclass[preprint,showpacs,preprintnumbers,amsmath,amssymb]{revtex4}

\usepackage{graphicx}
\usepackage{dcolumn}
\usepackage{subfigure}
\usepackage{bm}

\begin{document}

\preprint{APS/123-QED}

\title{Electrically Tunable Dicke Effect in Double-Ring Resonator}
\author{A. E. \c{C}etin}
\email{acetin@bu.edu} \affiliation{ Photonics Center, Boston
University, Boston, MA 02215, USA} \affiliation{ School of
Electrical and Computer Engineering, Boston University, Boston, MA
02215, USA}
\author{\"{O}. E. M\"{u}stecapl{\i}o\u{g}lu}
\affiliation{ Department of Physics, Ko\c{c} University, Sar{\i}yer,
\.Istanbul, 34450, Turkey} \affiliation{ Institute of Quantum
Electronics, ETH Zurich, 8093 Zurich, Switzerland}

\date{\today}

\begin{abstract}
The Dicke effect is examined in an all-optical system of an optical
waveguide side coupled to two interacting ring resonators in a
liquid crytal environment. The system is shown to exhibit all the
signatures of the Dicke effect under an active and reversible
control
   by an applied voltage.
\end{abstract}

\pacs{42.60.Da, 42.50.Nn, 42.79.Gn}


\maketitle


Substantial narrowing of the spectral line shapes due to collisions
of radiating and non-radiating atoms is first described by R. H.
Dicke in 1953 and called as the Dicke Effect \cite{Dicke53}. The
overall line shape consists of a superposition of a broad and a
narrow line shapes centered at the transition frequency. Such a
splitting of atomic decay into a pair of fast and slow decay
channels is closely related to the superradiance phenomenon
predicted shortly after the Dicke effect \cite{Dicke54}.
Superradiance is cooperative spontaneous emission of radiation from
an initially excited coherent ensemble of atoms. The slow and fast
decay channels are respectively named as subradiant and superradiant
decays of the system. Collective symmetric or anti-symmetric states
of an ensemble of atoms are respectively in superradiant or
subradiant phases.

In addition to the atomic ensembles, The Dicke effect has been
extensively studied in other systems, such as photonic crystals
\cite {photonic_xtal_dicke}, plasmonic lattices
\cite{plasmon_dicke}, electronic mesoscopic systems,
\cite{brandes,ApelPacheco08,OrellanaDiez06,OrellanaClaro04,PodivilovShapiro92,
VorrathBrandes03,brandes99,shahbazyan} and in electron waveguides
\cite{LeeReichl08,LeeReichl09}. Not all the signatures of the Dicke
effect can be found in many of these systems. Furthermore, control
of the Dicke effect is too challenging. Our purpose is to examine
the Dicke effect in an all-optical device with active and reversible
control. We find that a pair of microring resonators coupled to a
waveguide can exhibit all the key signatures of the Dicke effect in
a controllable way with the help of a nematic liquid crystal (NLC).
Tunable lifetimes of quasibound states in the resonators can be
translated into reversible active control of optical signals in
multiple ring resonator configurations. Such systems are used for
many optical communication and signal processing applications such
as all-optical logic gates \cite{ring_logic} and all-optical memory
elements  \cite{ring_memory}.

From a fundamental perspective, the system allows for controlled
investigations of quantum interference and decoherence by providing
an all-optical analog of an Anderson-Fano model which is the
prototype description of interaction between (quasi) bound and
(quasi) continuum states \cite{anderson,fano}. In addition,
extending the system to multiple ring configuration, quantum phase
transitions in the context of superradiance can be systematically
examined and probed (For a review see \cite{brandes} and references
therein).

\begin{figure}[t]
\centerline{
\includegraphics[width=8.3cm]{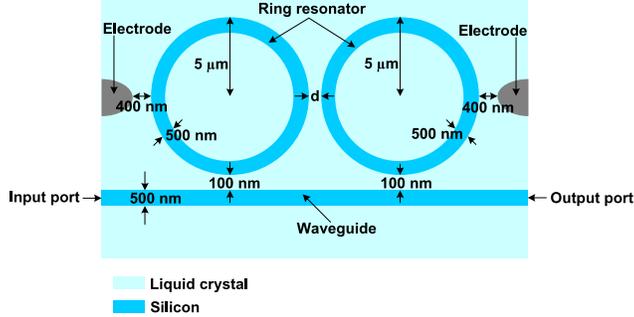}}
\caption{The dimension of the system consisting of two
ring-resonators on top of a waveguide feeded by two electrodes.}
\label{figure1}
\end{figure}

It has been shown that NLC can be used to tune the resonances of a
single ring resonator coupled to a waveguide \cite{MauneDalton03}.
To control linewidths, we introduce another ring to that system as
depicted in Fig. \ref{figure1}, where two identical microring
resonators at a distance $d$ apart are side coupled to a waveguide.
A TE polarized Gaussian beam (electric field is perpendicular to the
plane of the resonators) is sent from the input port of the
waveguide. Two electrodes provide a voltage to change orientation of
the NLC molecules used for cladding the resonators. Similar set up
but without NLC has been examined for its reflective properties
\cite{uzunoglu,dagli}. NLC allows for controllable coupling
coefficients between the resonators and the waveguide.

For a single ring, the resonance wavelength is determined by the
Fabry-Perot etalon resonance condition, $m\lambda_{m}=2\pi Rn_{eff}$
where $m=1,2,3...$, $\lambda_{m}$ is the wavelength of the mth
resonator mode, $R$ is the radius of the microring and $n_{eff}$ is
the effective refractive index for the waveguide mode
\cite{MauneDalton03,SalehTeich91}. Proximity coupling by the
evanescent tails
makes $n_{eff}$ depending on the refractive index of the NLC cladding,
$n_{clad}$ which is determined by

\begin{equation}
\frac{1}{n_{clad}^{2}}=\frac{\cos^{2}(\theta)}
{n_{e}^{2}}+\frac{\sin^{2}(\theta)}{n_{o}^{2}},
\label{eq1}
\end{equation}
where $n_{e}=1.744$ and $n_{o}=1.517$ at $589$ nm
\cite{MauneDalton03}, and $\theta$ is the angle of the NLC directors
(a local pseudo vector in the mean molecular long axis direction)
relative to the radial axis from the origin in the middle of the
electrodes \cite{MauneDalton03,Wiley98}.

\begin{figure}[t]
\centerline{
\includegraphics[width=8.3cm]{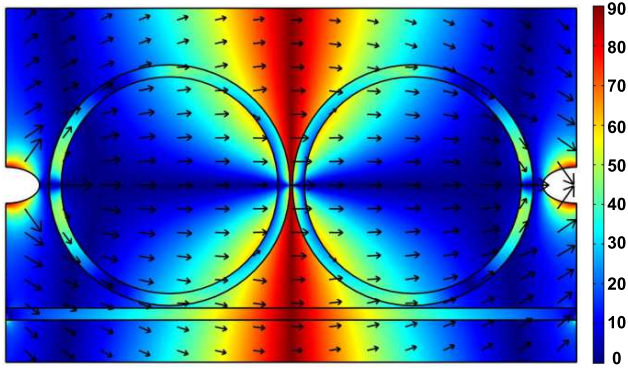}}
\caption{(Color) Angular difference $[^{o}]$ between the directors
of the NLC molecules and the radial axis. (Arrow) The direction of
the electric field generated by the electrodes at 5 V.}
\label{figure2}
\end{figure}

When there is no applied field, assuming the optical field is too
weak to induce any reorientation (optical Frederick's effect) of the
NLC directors, the NLC is in the isotropic phase, so that
$n_{clad}=1.596$. When sufficiently strong potential is applied from
the electrodes, the directors are deformed as shown in Fig.
\ref{figure2}. $\theta$ is locally determined by the Euler-Lagrange
equations, to minimize the free energy density of the NLC with given
elastic properties. For a potential field much stronger than the
elastic contribution, the directors will be fully polarized in the
applied electric field direction \cite{Collings02}. This permits
local modulation of the $n_{eff}$, analogous to electrooptic effect.
The electric field lines in Fig. \ref{figure2}, which is numerically
determined by the Poisson equation, indicate the director alignments
that can be controlled by the boundary conditions at the electrodes
\cite{MauneDalton03}. After $\theta$ distribution is found,
spatially inhomogeneous refractive index of NLC cladding is
calculated by Eq. \ref{eq1}, which is used to propagate the Gaussian
beam in the waveguide and to evaluate its transmission. We have
repeated this process for different potential values and different
separations between the rings.

Using transfer matrix formalism \cite{transfer_mat1,transfer_mat2},
reflection characteristics of coupled ring resonators have been
examined \cite{uzunoglu,dagli}. In order to examine the coupling
between the electromagnetic modes as realistically as possible, and
to treat NLC cladding correctly, we follow a computational approach
based upon finite element methods \cite{numerics}. We verify that
our numerical analysis yields the similar lineshape structures
obtained by the transfer matrix method. We calculate the resonances
by evaluating the intensity of the wave at the output port of the
waveguide, $I_{out}$. The spatially inhomogeneous $n_{clad}$ is used
in the coupled wave equations to solve for the resonator and the
waveguide modes and the evanescent waves in the NLC. Different
computational grids are used for every different geometry arising
when the ring separation is varied.

\begin{figure}[tbp]
\centerline{
\includegraphics[width=8.3cm]{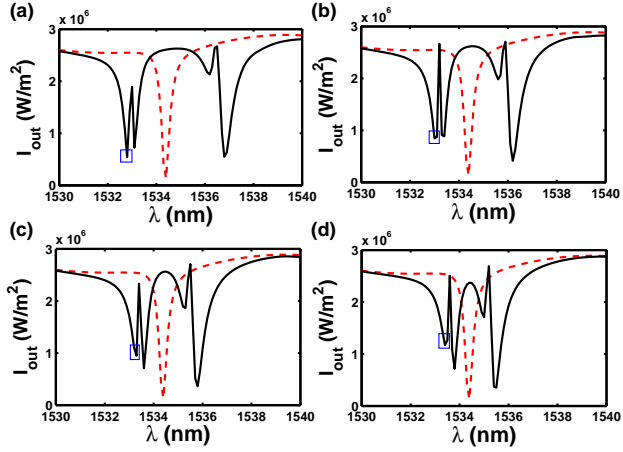}}
\caption{The resonance of the system as a function of wavelength,
for a single resonator (dashed line) and two resonators (solid line)
for different $d$ values, (a) 25 nm, (b) 50 nm, (c) 75 nm and (d)
100 nm at 0 V. } \label{figure3}
\end{figure}

The resonances of the system for different $d$ values without an
applied potential are shown in Fig. \ref{figure3}. Characteristic
splitting of the single ring resonance into four peaks is observed.
Four peaks arise due to simultaneous presence of direct proximity
coupling together with the additional bus waveguide mediated
coupling between the rings \cite{dagli}. In other systems such as
ballistic electron guides, the resonators are coupled only through
the bus waveguide and two peak splitting occurs. The asymmetry of
the lineshapes in Fig. \ref{figure3} is also a characteristic
signature of the quantum interference. The intereference is of Fano
type, due to spectrally overlapping subradiant and superradiant
decay channels. When the gap between two ring resonators is widen,
the proximity coupling between the rings reduce, while the bus-ring
coupling remains unchanged. The splitting between the symmetrically
placed symmetric and antisymmetric modes about the isolated ring
resonance decreases. The further splitting of these modes due to the
ring-bus interaction is independent of $d$. On the other hand, their
width ($\Gamma$, Full-Width Half-Maximum (FWHM)) varies with the gap
between the rings.

In addition to the splitting into subradiant and superradiant
channels, we have found the oscillatory behavior of the linewidths
with the distance between the resonators. This signature of the
Dicke effect is due to the spatial interference of radiation from
decaying quasibound states of the rings coupled to the waveguide at
separated locations, inherent to the collective nature of the
system. Dicke oscillations of symmetric and antisymmetric modes are
translated into further split modes in our case. The effect can be
understood analogous to the level repulsion between coherently
coupled degenerate bound states \cite{LeeReichl09}. Together with
the interaction caused splitting of the resonances, interacting
(interfering) decay channels also split. Representing linewidths as
imaginary parts of the resonances, the symmetric and antisymmetric
modes lie $180^o$ out of phase in the complex plane. The interaction
channel through the bus waveguide introduces $d$ dependent phase
accumulation to the mode freely propagating in between the rings
along the waveguide. This is translated as a sinusoidal coupling
between the rings or a rotation operation in the complex energy
plane. By increasing the gap between the rings, the resonances
collapse back onto the isolated ring value in a spiraling fashion.
During this spiral motion, their imaginary parts periodically become
vanishing and finite and hence oscillates.

\begin{figure}[tbp]
\centerline{
\includegraphics[width=8.3cm]{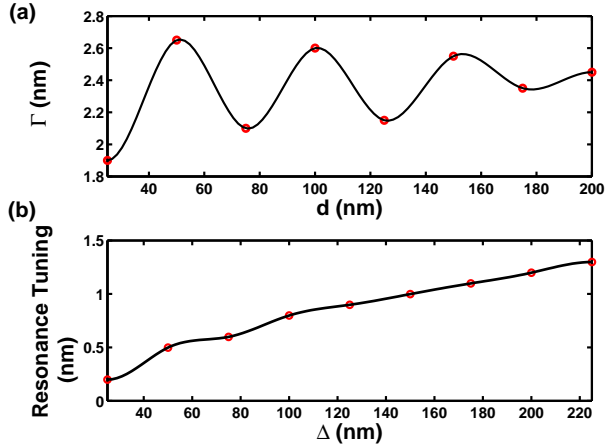}}
\caption{(a) Width of the chosen resonance in Fig. \ref{figure3}(a)
vs the gap between two ring resonators (b) resonance tuning vs
additional gap distance to $d=25$ nm at 0 V.} \label{figure4}
\end{figure}

Choosing a resonance, the most energetic one, labeled with a blue
box as in Fig. \ref{figure3}(a) at $d=25$ nm, we investigate the
dependence of the resonance width and resonance wavelength on the
spatial separation between the rings. Fig. \ref{figure4}(a) shows
the variation in the width of the resonance as a function of the gap
between two ring resonators. The coupling between the rings
decreases with distance and the oscillations eventually disappear.
The width saturates at the single ring value at long distances. In
one-dimensional system of bi-ripple ballistic electron waveguide,
sinusoidal periodic behavior of the resonance width is found as the
coupling does not decay with the distance \cite{LeeReichl09}. Our
situation is similar to traditional Dicke systems in higher
dimensions, where the coupling decreases with the distance
\cite{brandes99,shahbazyan}. Approach to the single ring resonance
(resonance tuning) as we increase $d$ by an amount $\Delta$ is shown
in Fig. \ref{figure4}(b). As the coupling decreases with distance,
so does the splitting of the collective modes, and thus, the most
energetic mode becomes less and less different from the single ring
resonance.

\begin{figure}[tbp]
\centerline{
\includegraphics[width=8.3cm]{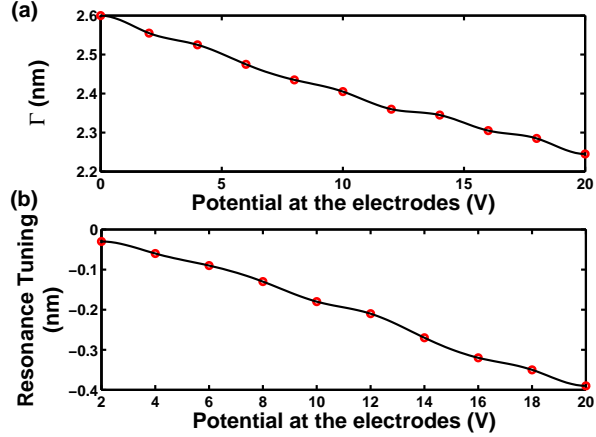}}
\caption{(a) Resonance width and (b) resonance tuning of the chosen
resonance in Fig. \ref{figure3}(d) at $d=100$ nm vs applied voltage
from the electrodes.} \label{figure5}
\end{figure}

Considering now another most energetic resonance found at a larger
separation between the rings, labeled with a blue box as in Fig.
\ref{figure3}(d) at $d=100$ nm, we examine possible active control
of these signatures of the Dicke effect demonstrated in Fig.
\ref{figure4}. When the potential applied from the electrodes is
turned on, the linewidth and the resonance are changed as shown in
Fig. \ref{figure5}. The linewidth decreases with increasing
potential, which aligns the directors such that in the coupler
zones, the cladding index increases to $n_e$. This locally reduces
the index contrast with the silicon guides and enhances the field
penetration, and thus the proximity coupling between the resonators
increases. The chosen most energetic mode becomes further split from
the single ring resonance with the increasing coupling coefficient.
The applied voltage and the spatial separation between the rings
have opposite effects on the coupling resonators.

\begin{figure}[t]
\centerline{
\includegraphics[width=8.3cm]{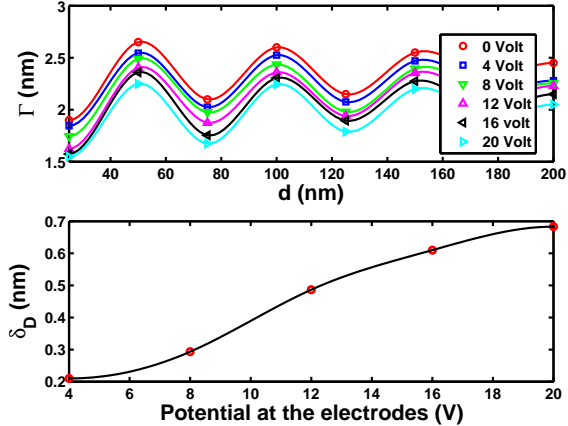}}
\caption{(a) Width of the chosen resonance in Fig. \ref{figure3}(a)
vs the gap between two ring resonators for different voltage values
applied from the electrodes (b) Increase in the periodicity of the
resonance width function for different voltage values.}
\label{figure6}
\end{figure}

Finally, we analyze the influence of the external potential on the
periodicity of $\Gamma$ as a function $d$. Fig. \ref{figure6}(a) is
in agreement with the earlier observations in Fig. \ref{figure5}(a)
that the resonance width decreases with the applied voltage. The
periodicity of  $\Gamma$ when no voltage is applied (see Fig.
\ref{figure3}(a)) is determined to be $\lambda_{D}=48.92$ nm. Change
of the period from this value when the potential is applied is
denoted by $\delta_{D}$. Fig. \ref{figure6}(b), shows that the
periodicity of the $\Gamma$ increases steadily with the potential
applied from the electrodes. Enhanced coupling of the rings
increases the splitting of symmetric and antisymmetric modes. For
the chosen most energetic resonance, this makes it closer to the
waveguide mode. Thus the free propagation and the associated phase
accumulation between the rings occurs at smaller frequency or at
larger period. In the complex energy plane, that means the rate of
rotation with $d$ is reduced.

Summarizing, we have examined the resonances and their widths in an
all-optical system of a pair of ring resonators side-coupled to an
optical waveguide in a NLC environment. We have found that the
system exhibits all the key signatures of the Dicke effect,
splitting of the lifetimes into slow and fast decaying channels in
an oscillatory manner with the separation of the resonators. The
energies and the lifetimes of the quasibound states of the
resonators can be controlled with the applied voltage, which allows
for tunable longer range interactions between the resonators. Such a
reversible active control can be used to examine coherent collective
effects, decoherence, and superradiant phase transitions. Besides,
fine tuning of finesse and full spectral range can be exploited for
multi-ring systems, such as CROW \cite{crow} and SCISSOR
\cite{scissor} configurations, as well as rings coupled to multiple
waveguides. These systems are subject to many applications,
particularly optical logic and memory, filtering, reflectivity and
optical switching.


The authors acknowledge useful comments by A. Serpenguzel, H. Altug and H. Tureci.


\bibliography{references}

\end{document}